# Human Task Monitoring and Contextual Analysis for Domain Specific Business Processes

Kunal Suri, Adrian Mos

Xerox Research Centre Europe
6 Chemin de Maupertuis, 38240 Meylan, France

{kunal.suri, adrian.mos}@xrce.xerox.com

**Abstract.** Monitoring the execution of business processes and activities composing them is an essential capability of Business Process Management (BPM) Suites. Human tasks are a particular type of business activities, and the understanding of their execution is essential in effectively managing both the processes and human resources. This paper proposes a monitoring framework with a capability to monitor and analyze the human tasks in a domain specific setting and contextually correlate the task execution patterns to the workload distribution on human users.

The framework uses the notion of concept probes that match the business concepts used in definition of business processes. The proposed human task monitoring and contextual analysis (HTMCA) component considers multiple artifacts involved in the execution of a human task, rather than focusing only on classic activity/task metrics retrieved from BPM engines.

This approach aspires to provide two main advantages to organizations using it. Firstly, it enhances the understanding of the workload of human users that participate in people-intensive business processes under various roles. Secondly, it gives organizations tools and insight for fine-tuning their user performance taking into account the specific context of their business various artifacts domains.

The proposed framework builds on previous work that lays the basis of vendor-independent, concept-centric BPM monitoring, and provides the critical missing element of human task understanding. This has the potential to significantly benefit any BPM deployment and the validation work is in advanced stages of building a full prototype that demonstrates this value in a realistic industrial setting.

# 1  INTRODUCTION

Business Process Management (BPM) is an important paradigm for today's enterprise solutions as it provides a level of agility to businesses by enabling them to quickly adapt to the changing business requirements and customer needs. BPM Suites are complex software stacks that execute business processes (BPs) and connect them to various enterprise resources including services deployed using Service Oriented Architecture (SOA) design patterns or legacy applications. In general, a BP is a systematic aggregation of activities which includes user activities or service/automated activities, all of which execute under a defined time frame to provide a certain value to the end user. Any enterprise that aspires to maintain its expected business value from its business processes should perform continuous monitoring of BPs and compare the output to the pre-defined key performance indicators (KPI's) to check deviations in the expected process output.

In classical business processes monitoring there is not much focus on business domains and business semantics which is envisioned to be resolved by introducing a domain specific monitoring framework where such business semantics could be introduced easily. In other words, a domain specific BPM running a process having 'Payment' concept, wherein the 'Payment' concept consists of one or more activities and SOA based web services executed systematically, would have the business understanding of 'Payment' concept and it relation to different departments and geographies. Such a framework would also enable better governance and SLA management of processes by making use of business concepts which would allow the organization to be agile to market dynamics. This agility would be provided by enabling mechanisms to propagate changes to a specific business concept that would instantaneously be realized in all processes using that concept. For e.g. 'document scanning', 'handling tax calculation' (country specific), 'discharging of a patient in a hospital', all of them can be specified as an atomic business concept by a domain experts for a specific domain like BPO, finance and healthcare. Thus an atomic business concept in Finance could be visualized as a set of activity and SOA services executing together to provide a business value. A normal domain specific process would consist of various atomic business concepts that would have pre-defined business semantic knowledge about the business concepts and the set of the activities and SOA or other layers involved in it. This would mean to introduce a change in taxation policy we just need to change the 'handling tax calculation' business concept and all the processes in an organization utilizing this concept would realize that changes without the need of going to each processes and introduction these changes.

Furthermore, this paper builds on previous work done on domain specific monitoring [1] which defined a mechanism to perform aggregated monitoring of the activities running in the BPM suite and the corresponding web services running in the underlying SOA layer. But in this paper we keep our focus on introducing the vision and capability to monitor and analyze human tasks in a domain specific BP environment rather than restating the basic the mechanisms of domain specific business process monitoring.

The proposed monitoring of the execution of human tasks[1] in domain specific business processes is of great value because it provides an understanding of their execution which is essential in effective management of both the processes and human resources that in some domains (e.g. healthcare, finance) are very expensive and scarce. This framework would help in aggregating data and generating the related knowledge required by both technical and business experts to get better insights about human tasks execution. This knowledge can aid re-engineering of BPs by updating task priorities, efficient users assignment, creating new roles for complex tasks that require domain knowledge etc. Lastly, the system allows business process owners to specify constraints and alerts on human tasks involved in a particular business concept. These alerts would be implemented and propagated simultaneously across all the deployed business processes utilizing those concepts.

## 1.1  Motivation and Research Problem

Business processes are based on certain business needs expressed as business concepts in a business domain, requiring them to orchestrate various automation services and/or human resources in a tightly coupled manner to generate a value for the end customers. This implies that monitoring each component used by a BPM without understanding the domain and deployment semantics i.e. as isolated components would not be very useful for locating the BP's weakness. The generic monitoring of individual layers used for BP execution creates gaps due to unavailability of semantics about the business domain in which the processes are deployed, resulting in various crucial questions not being answered, some of them being:

- Which layers should be monitored to get a holistic view of the process and business concept it relates to?
- How to pinpoint the layers and the components that cause bottlenecks during a business process execution?

---

[1] Human task/ activities are task/ activities involving human actors that perform the task with the assistance of a software application like BPMS or under supervision of a computer/software application, captured though some device or sensor. User task defined in BPMN 2.0 are one type of human task.

These issues with generic monitoring of individual layers like BPM, SOA, Network Layer etc. involved in a BP execution motivated the vision of a domain specific monitoring framework [1] which presented a multi-level cross-layered monitoring of various components involved in execution of the business processes. It also gave forth the idea of understanding the domain specific business semantics by using concept probe which lays the basis for vendor-independent, concept-centric BPM monitoring.

The above mentioned work proposed in [1] somehow missed to focus on activities performed by human actors in more depth. Like in any other classical monitoring techniques, the human tasks were handled just like any other ordinary tasks by retrieving the related metrics from the BPM engine. This led to some of the questions related to human resource engagement in a BPM to be unanswered, such as:

- How to visualize under or over utilization of humans involved in BPs and help in business re-engineering efforts?
- Which factors are generally responsible for Service Level Agreement (SLA) violation in context to human tasks?

These questions motivated us to introduce a separate layer for human task monitoring fully integrated in the domain specific business process monitoring framework proposed in [1]. The HTMCA layer helps domain experts for defining various metrics needed for analyzing human related tasks in any domain specific business process setting.

Furthermore, it is evident that there can be various extrinsic factors causing the deviation in execution of human tasks such as weather, geographical location etc. These factors though important fall in the category of environmental, behavioral or social sides of human – computer interactions and may not be easily captured by a BPM and other layers. Thus we have not touched these factors in our work and this paper focuses mainly on time complexity and workload i.e. number of task in user's worklist to perform an analysis.

The rest of this paper is structured as follows. Section 2 presents the overview of the approach. Section 3 presents the structure of HTMCA layer. In Section 4 we present some experimentation details. We discuss related work in Section 5. Section 6 summarizes and concludes the paper along with outlining the future work.

## 1.2 Context Analysis using HTMCA

**Fig. 1** illustrates one of the possible task execution analysis that can be performed by using HTMCA for a task executed by a user x, for task 'Aα' using concept probe 'CPα'. The figure depicts that the user x, at the time of execution of the task 'Aα', was involved in multiple activities. Some of these tasks were running before the start of task 'Aα' and some tasks were started during the execution of 'Aα'. It correlated all this data to create knowledge for understanding the violations in human task execution and reasons for delays.
The
**Fig. 1** depicts the execution of task 'Aα' by user x, during which the user completed task T2 that was started in the same time interval 'tα'. The user x also managed to complete task 'T3' that was started before the start of task 'Aα'. The user also had to work on execution of task 'T4 and Task 'T5'.
Such analysis could be very useful to understand the human workload in case of human intensive processes like workload on a Nurse in healthcare processes, and enable the experts to visualize the involvement of this nurse with various processes for various patients in a hospital.

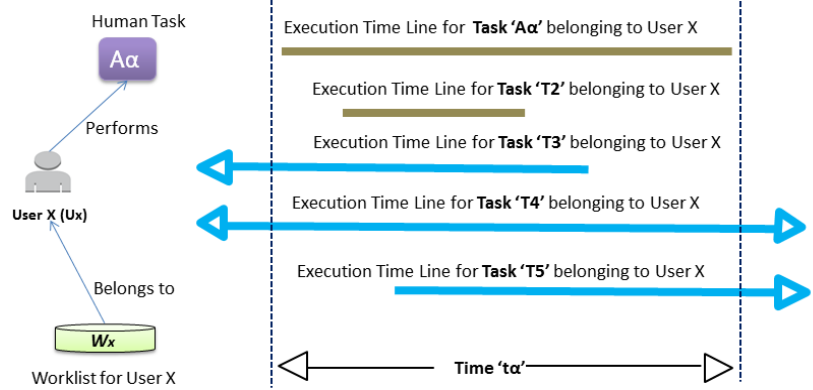

**Fig. 1.** Workload Pattern of a Human Actor in a Business Process

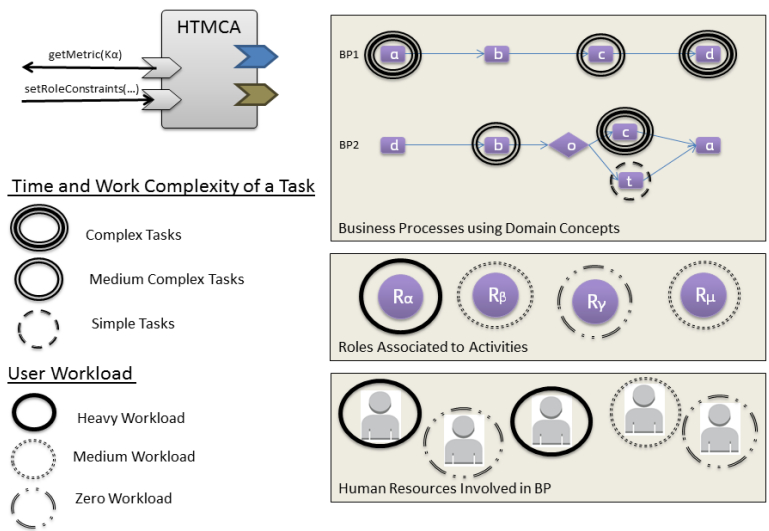

**Fig. 2.** A Conceptual View of Human Workload Analysis

**Fig. 2** depicts a conceptual view of work complexity and workload on various users involved in different business processes using HTMCA to analyses the human tasks. This figure provides one of the many possible analyses that can be performed using HTMCA. BP1 and BP2 are business processes running in a Domain Specific BPM for domains like Finance or healthcare. The HTMCA would have the semantics of task complexity provided by a domain expert in a domain specific setting for performing a task with minimal errors. These tasks vary between complex, medium complex or simple task. The workload on a user at any given point of time is calculated by correlating the number of roles belonging to a user and the number of tasks that are active and pending in user's worklist. Thus, at any instance a user can be working under heavy, medium and zero workload. Furthermore, domain experts would be used to outline the definition of task complexity and its workload measurements during the creation of business processes. Such analysis provides an improved reporting mechanism to the business owners and the leadership, allowing them to initiate the business process re-engineering actions.

## 2   OVERVIEW OF THE APPROACH

This paper extends the domain specific monitoring framework of Mos [1] by introducing human task monitoring capabilities for all the user activities that are involved in a domain specific business process. The approach entails the extension of Concept Probes (CPs) that are monitoring entities corresponding directly to business concepts and provide the aggregated information from various layers involved in the execution of Business Process. The creation and binding of CPs is similar to [1] and thus not discussed in details in this paper. In addition to extending the CPs, this approach also extends the notion of Business Process Probe (BPP), which corresponds to each deployed BPs and is essentially a composition of the CPs that corresponds to the concepts used in the business.

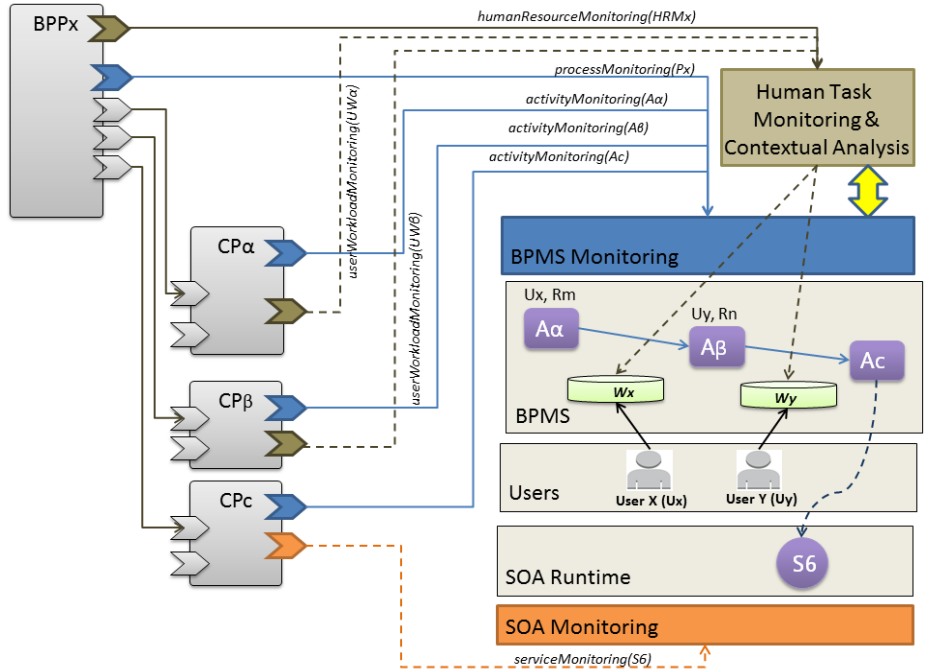

**Fig. 3.** Different Components Involved in Domain Specific Monitoring of BP

The **Fig. 3** shows the activity 'Ac' connected to the SOA service 'S6' and user activities 'Aα' and 'Aβ' performed by Role 'Rm' & 'Rn' assigned to User 'Ux' and 'Uy' respectively. There exists a User (U), Role (R) and Worklist (W) for each human activity that is involved in a BP. For simplicity reasons, we have depicted that each human activity such as 'Aα' that is assigned a Role 'Rm' has a single user 'Ux' but we are aware that a single role can be assigned to multiple users and also a single user can have multiple unique roles based on need of business. The SOA links represent regular web service calls such as SOAP or RESTful invocations.

The **Fig. 3** also presents three concept probes 'CPα', 'CPβ' and 'CPc' that corresponds to the business concepts α, β, c, (Such as *Payment* for *Finance* or *Patient Discharge Checkup* in *Healthcare*) used in the illustrated BP through the activities 'Aα', 'Aβ' and 'Ac'. In addition to the CP's, the figure also shows a business process probe, the 'BPPx', which corresponds to the example BP and uses the three CPs i.e. Two CP's for human tasks and One CP for automated task to aggregate BP-level information. The outgoing lines from the CPs represent their connections to the BPMS, HTMCA and SOA monitoring systems.

   For example, since 'CPc' is a probe specifically generated for service/automated 'concept c', it will interrogate the BPMS monitoring system regarding the activity 'Ac' along with the SOA monitoring system regarding services 'S6'. These connections are generated based on prior knowledge that 'concept c' is used for business activity Ac and its underlying web services S6. This knowledge comes from mapping the concept probes to the business concepts. The lines are labeled with abstract functions that simply illustrate what kind of data they collect from the monitoring systems. Similarly the 'CPα' is a concept probe specifically generated for human task probe 'concept α' and 'CPβ' for human task probe 'concept β'. These CP's would interrogate the BPMS to get the information about activity execution and the HTMCA to get the human-centric contextual information.

The CPs leverages the aggregated data by performing monitoring of multiple layers involved in BP execution i.e., BPMS, HTMCA and SOA systems. The CP's can further include incoming data from operating system (O/S), data from cloud infrastructure like server location and other network information depending on the application deployment. It is to be noted that the information provided by a BPP is significantly richer than that provided by BPMS monitoring systems for an entire BP because it includes the breakdown of monitoring information for each concepts used in the BP as well as the aggregated BP-level knowledge. Naturally, modern BPMS monitoring systems can make the correlation between a BP and its composing activities but our approach consolidates monitoring information in a conceptual manner from various layers along with the semantics business concepts.

## 3 HUMAN TASK MONITORING AND CONTEXTUAL ANALYSIS STRUCTURE AND FUNCTIONALITY

The **Fig. 4** presents the structure of the HTMCA, which contains three main components. First component is called the 'workload data collector', which collects raw data about the worklist, roles, users and execution details of the activities from the BPMS. The second component, 'workload analysis' aggregates raw data obtained from the collector component into composite metrics. These composite metrics are data structures that present the aggregate monitoring information combining the individual metric data for the BPMS and user workload. The third component, 'workload alerts and reporting' relates to the ability of the HTMCA to provide specific data reports about the execution of human activity and register rules for alerting. This component also allows the registration of SLA, setting constraints on various roles through the 'configure alerts' port and uses the 'workload analysis' component to constantly compare the aggregated metric values with the required thresholds. If SLA thresholds are exceeded it can notify registered 'monitoring listeners'. These listeners are external entities (out of scope of this paper), which can be connected to the monitoring probe and notified of important alerts and events.

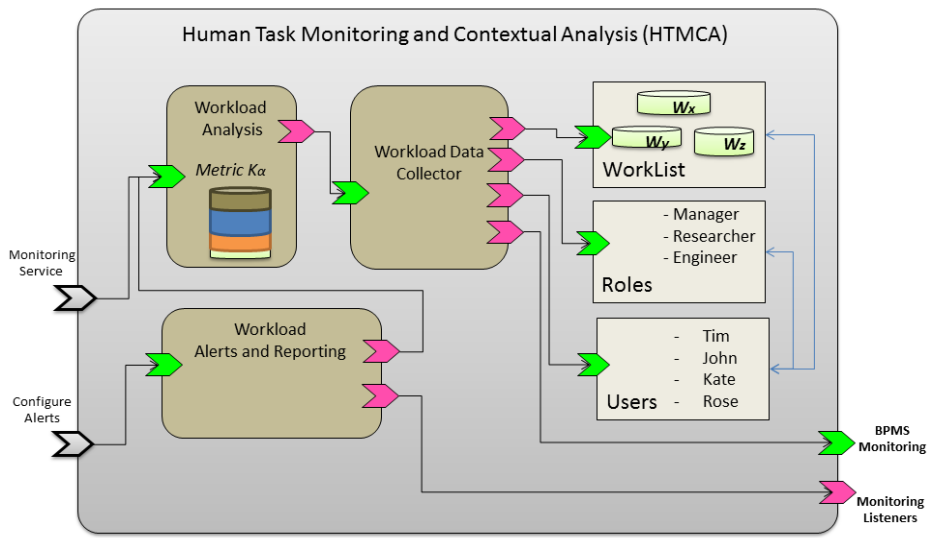

**Fig. 4.** HTMCA Structure

## 4 EXPERIMENTATION

The implementation of the HTMCA layer along with concept probe is technology independent. We have created a proof-of-concept for our research on HTMCA and the concept probes' using open source tools such as, Eclipse based Stardust BPM 2.1.1 tool [2], which is an open-source, fully featured BPM environment provided by SunGard. The SOA environment is provided by JBoss Fuse ESB 6.1.0 [3] from Red Hat as depicted in **Fig. 5**. The code developed for HTMCA and the concept probes has been fully developed in Java. The HTMCA data can be exposed using JMX MBeans which could be read by any client subscribed to these MBeans.

In our experimentation, we make use of the API exposed by the Stardust BPM to retrieve the relevant information from the Audit Trail Database (Derby) about the completed business processes and activities. We also make use of the API based on event generation in Stardust for capturing data about changes in state of tasks. This data is used by the HTMCA component to perform the co-relations and analysis of the user workload with respect to ongoing tasks. The monitoring framework also provides the possibility to provide different views for analyzed data to answer various research problems for human task monitoring, thus generating useful knowledge for business leadership. This data can be further linked to various reporting system for warehousing and deeper understanding.

We have developed simple consoles that help in visualizing data from HTMCA about the user workload. This data can also be fed to Eclipse based graphical editors for displaying the data along with the graphical process model.

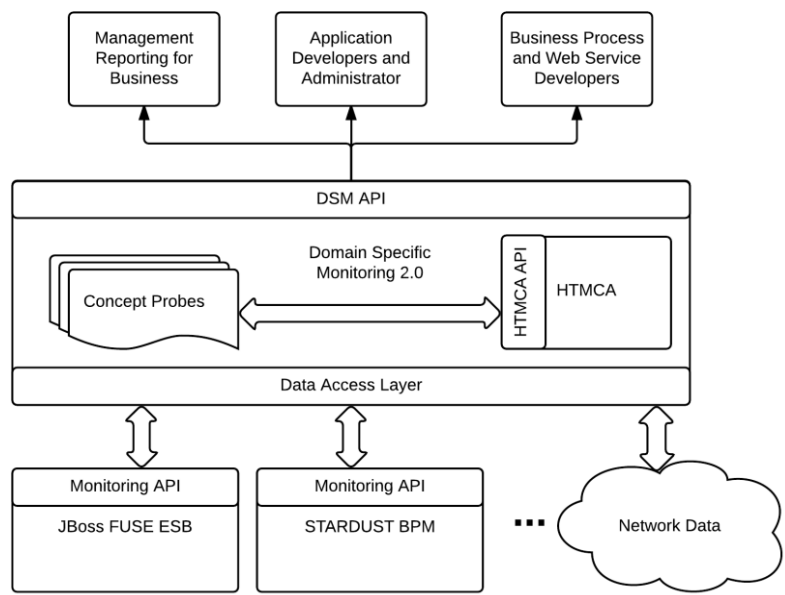

**Fig. 5.** Setup for Domain Specific Monitoring Using HTMCA

Some functionality provided by HTMCA API for retrieving data related to human workload is explained as follows:

*getAllRolesGrantedToUser()*: This function takes the username as input and provides a list of roles a user can perform i.e. all roles that have been assigned to a user in question. This function can be used at any point of time to review the role-based load on any user with respect to completed, active or pending human tasks. It gives insight for finding over utilization and underutilization of human resources.

*getAllBPExecutedbyThisUser()*: This function takes the username and time range such as, hour, day, week or months as input and retrieves the details of all the business processes that were executed by a particular user in that specific time frame. It provides a detailed overview of all the Business Process that a particular user worked on, listing each activity in a business process during that specified time the user worked on along with activity execution information like execution time, activity id, activity start time, activity end time, role under which the activity was performed etc.

*getAllBPModelGrantedToUser():* This function takes the username as input and provides entire list of Business Processes that are granted to this user. In other words, it provides a list of business processes a user is allowed to work on. It is useful to retrieve user related details from a high level i.e. at business process level.

*getActiveActivityDetailsUserRelated():* This function takes the username as input and provides a detailed list of all the activities that are present in the worklist of a specific user and have still not being completed. This function provides a correlation to the number of tasks active in a user's worklist, the amount of time those tasks have been active and depending on the priority of pending task which tasks have an alert mechanism to alert the user about SLA associated with the task. Such complex functions provide business process stakeholders tools to understand the work pattern and load on any user at any given time.

## 5  RELATED WORK

There are several commercial tools and academic approaches to monitor business processes and services in a BPM/SOA setting, but most of them are generic from the perspective of business domains and they do not focus on providing a distinct monitoring technique for human related tasks. An in-depth discussion of such generic BPM/SOA monitoring work can be found in [1]. Many good research techniques in the area of business processes monitoring and SOA monitoring, consider monitoring these components individually and not giving much consideration to aggregated monitoring of these layers. Furthermore, these research techniques do not focus much on understanding the effect of human workloads and its correlation to violations of SLA or other issues.

   WS-Human Task specification for BPEL4People [4] [5] [6] [7] provides extension to BPEL for providing it a better capability for supporting a broad range of scenarios involving human resource in a business process. This specification promoted by various organizations such as IBM, SAP was a step towards standardization for handling one of the most important resources i.e. humans involved in a BP. It talks about various aspects of human activity management involving issues such as delegations, escalations etc. Certain work done in terms of human resource management like [8] by Kumar et al. discuss about the mechanisms to distribute work dynamically by using strategies involving security and performance considerations rather than the generic pull and push approaches applied to human workload distribution as employed by normal workflow systems. These approaches talk about workload distribution but do not focus much on the monitoring framework that can provide insights to use these distribution mechanisms. Such approaches have a potential to be bundled on top of our HTMCA for distributing workload automatically and in a more efficient manner. Rhee et al. [9] describe automatic rational task allocation and work-item prioritization techniques to provide performance efficiency to a generic user-centric business process by combining the process engine perspective and task performer perspective to attain a better task allocation mechanism. They differ from our approach, as they do not consider the contextual monitoring of user-centric business processes deployed in domain-specific settings. Such work have a potential of being complimented with HTMCA for adaptation mechanism by creating efficient and automated distribution rules for tasks based on the monitoring output. Ha et al. [10] have suggested methods to transform business processes into queuing network models, in which the agents are considered as servers. Calculating workloads from server utilization perspective and determining task assignment policy to balance workloads. Such techniques tend to ignore the performance effect of long periods of workload on users and the nature of humans which creates issues with multitasking and the fact that human computing cannot be expanded in a way similar to servers.

   There have been discussions on avoiding violation to SLA's involved in a business process by using improved mechanisms of handling service activities by means of advanced web service automation techniques but it still doesn't cover the gaps leading to SLA violations by mismanagement of human resources. Recent research on SOA layer provides various mechanisms for BP improvement as discussed in [11] Risk-Mitigation Framework for Business Transaction at Run-Time involving active monitoring of the generic business transactions, computing the potential risks and providing adaptation to services in a service-based application to avoid Global SLA violation. An understanding of web service signature and protocol incompatibility and ways of automatic service adaptation [12][13][14], estimation remote web service execution times [15] and advanced Service Oriented Architecture monitoring involving end-to-end monitoring framework for SOA [16] by focusing on monitoring of the integration layer responsible for mediation, routing, transporting service requests, puts forward mechanisms that could reduce SLA violations in SOA layer and thus in BPM using it. Yang et al [17] discuss on support for human task execution in context of service compositions by providing scheduling algorithm to find better performance human resources. Rodriguez et al [18] provide a description and importance of human task management in Healthcare using BPM. Moldovan et al [19] talk about cross-layer multi-level monitoring for analyzing cloud services that could be used complementary to the work done on HTMCA and concept probes running on cloud architecture for specific businesses.

Therefore, in contrast to the above-mentioned work that treats humans as any other server resource, the HTMCA layer is a novel contribution for domain-specific analysis of human tasks in context of workload and human roles.

## 6  SUMMARY AND CONCLUSION

The framework described in this paper proposes a human task monitoring and contextual analysis (HTMCA) layer, which correlates various artifacts involved in the execution of a human task, rather than just focusing only on classic task metrics retrieved from BPM engines. The paper builds on previous work that lays the basis of vendor-independent, concept-centric BPM monitoring of automated tasks. It provides a better insight for understating human task mapped to business concepts executed in BPM Suite. These insights are crucial for proper management of scarce and expensive human resources in various domains such as healthcare, financial services to name a few.

For technical users, application developers and application owners, the domain specific monitoring framework provides ability to breakdown performance problems from each layer i.e. service, human or cloud. The business owners would get a better understanding of the human-resource usage and its impact, along with knowledge of various other automation layers involved in fulfilling an end-to-end process.

Having business concepts at its core, the framework aggregates cross-layer data from multiple components such as BPM, human layer and SOA environments in order to provide a holistic view of the applications execution state. This enables contextual correlation of tasks, human actors, roles and the execution metrics of activities from BPM & SOA, which is helpful for further in-depth analysis of SLA violations.

A full prototype of the presented framework on HTMCA and Concept probes is in advanced stages of implementation, using Stardust BPMS [2] and JBoss Fuse ESB [3] as the target BPM and SOA layers, respectively. The implemented probes correspond to the concepts of a sample domain chosen for validation correlating the data for BP human activity execution with HTMCA. The prototype also consists of a simple console showing the collected data that can also be fed to Eclipse-based graphical editors for displaying data within graphical process modelling environments.